\def\e{\begin{equation}}
\def\ee{\end{equation}}
\def\ea{\begin{eqnarray}}
\def\eea{\end{eqnarray}}
\def\nm{\nonumber \\ }
\def\bc{$WBC_2 \ $}
\def\v#1{\vert #1 \rangle}

\def\NPB#1#2#3{{\sl Nucl. Phys.} {\bf B#1} (#2) #3}
\def\PLB#1#2#3{{\sl Phys. Lett.} {\bf #1B} (#2) #3}

\def\IJMPA#1#2#3{{\sl Int. J. Mod. Phys.} {\bf A#1} (#2) #3}
\def\LMP#1#2#3{{\sl Letters in Math. Phys.} {\bf #1} (#2) #3}

\documentstyle[12pt]{article}
\advance \topmargin by -\headheight
\advance \topmargin by -\headsep

\evensidemargin \oddsidemargin
\begin{document}

\title{ \begin{flushleft}
\normalsize{ ISSN 0133-462X \hfill ITP Budapest Report No.503 }
  \end{flushleft}
\vspace{5cm}
 Null vectors of the $WBC_2$ algebra }
\author{\normalsize{ Z. Bajnok\thanks{ E-mail: bajnok@ludens.elte.hu}}
 \\ \\
 \normalsize{ Institute for Theoretical Physics}\\
 \normalsize{   Roland E\"otv\"os   University} \\
 \normalsize{ H-1088 Budapest, Puskin u. 5-7, Hungary} }

 \maketitle

\vspace{1.5cm}

 \begin{abstract}

Using the fusion principle of Bauer et al. we give explicit
expressions for some null vectors in the highest weight
representations of the \bc  algebra in two different forms.
These null vectors are the generalization of the Virasoro ones
described by Benoit and Saint-Aubin and analogues of
the $W_3$ ones constructed by Bowcock and Watts. We find
connection between quantum Toda models and the fusion method.

\end{abstract}
\newpage

Conformal field theories describe the statistical mechanical
systems at their critical point. Their infinite dimensional
symmetry algebra, the Virasoro algebra or its extension,
plays an important role in analyzing them. In most cases these
models correspond to degenerate representations of these
algebras. Such representations are crucial since on one hand
they possess a closed operator algebra with only a finite number
of primary fields, while on the other hand they contain null vectors.
These null vectors lead to differential equations for the
correlation functions, and in this sense, they can be regarded
as equations of motion governing the dynamics of the theory.
Although this application needs explicit expressions for
the null vectors such formulas were not given  in general until
the last few years.

Recently there has been great interest to describe
these null vectors [1--6].
First Benoit and Saint Aubin constructed very compact
expressions for a subclass of the Virasoro null vectors then
Bauer et al. rewrote this formula using the fusion method.
They showed how to produce null vectors in the generic case
by fusing explicitly known null vectors with primary fields.
Later Bowcock and Watts extended this method to $W$ algebras.
They found that the fusions of the null vectors and of
their descendants should be investigated simultaneously.

In this paper we use their method to produce null vectors in the
\bc algebra. Explicit expressions for the simplest null vectors
are given. In the fusion method we deal with only those primary
fields which belong to completely degenerate representations and
have two independent null vectors at level one and two. It turns
out that there is a special null vector among the descendants ones.
This null vector is nothing but the quantum equation of motion of
the appropriate Toda model and in order to produce null vectors
one has to study the fusion of this descendant null vector.
We start with the representation theory of the \bc algebra.
Then we describe how to handle the action of a spin $j$
chiral current in the fusion procedure. Finally, using this result we
study two different cases namely the case of the $B_2$ and
$C_2$ Toda models, respectively.

The \bc algebra is one of the simplest $W$ algebras. It is
an extension of the Virasoro algebra generated by the
Virasoro field, $L(z)$, and a spin 4 chiral current, $W(z)$
\cite {BC2}. The highest weight (hw.) representation of the
algebra contains a hw. vector, $ \v{hw} $, which satisfies:
\e
 L_n\v{hw}=\delta_{n,0} h \v{hw} \quad ; \quad  W_n\v{hw}
 =\delta_{n,0}w \v{hw} \quad ; \quad  n\geq 0
\ee
In order to analyze the determinant formula \cite{KaWa} we use
the following reparametrisation of the $W$ weights :
\ea
&h(x,y)={Q \over 4}(x^2+2xy+2y^2)-{1\over 4}(5Q+{10\over Q}-14) \nm
& w(x,y)=\bar B \biggl(
     4Q(Q-3)(27Q-32)y^3(y+2x)-Q(3Q-2)(16Q-27)x^3(x+4y)\nm
   &\qquad +{(Q^2-2)\over
Q}\Bigl(14Q(x^2+2xy+2y^2-6Qx^2y^2)+Q-6+{2\over Q}
   \Bigr)  \biggr)
\eea
where
$$
\bar B^{-1}={32\over Q^2}
{\sqrt{(4Q-5)(5Q-8)(7Q-6)(3Q-7)(3Q-2)(Q-3)(75Q^2-226Q+150)}}
$$
Here we used $Q$ instead of the central charge (C). They are related
to each other as \break $C=86-60/Q-30Q $.
Now the determinant formula dictates that if
\e
x=a-c\  {2\over Q} \quad {\rm or} \quad  y=b-d\ {1\over Q}
\ee
then the representation is degenerate and there is a null vector
level $ac$ or $bd$. If both conditions hold at the same time
then the representation is completely degenerate and contains
two independent null vectors. We restrict ourselves to only
such representations. The hw. vector in this case is denoted by
$ \v{ab;cd} $ and the two independent null vectors by
${\cal O}_{a;c}\v{ab;cd}$ and ${\cal O}_{b;d}\v{ab;cd}$,
respectively.

To produce null vectors we follow the method of Bauer et al..
They considered the fusion of a primary field, $\phi_{h_0w_0}(z)$,
belonging to a degenerate representation, with another primary
field, $\v{hw} =\phi_{hw}(0)$. Setting the null vector equal to
zero and demanding its consistency implies recursive equations for
the descendant fields appearing in the OPE. Solving these equations
iteratively explicit expression can be obtained for the null states
in the corresponding representations. Everything one needs to know is
the action of generators after the fusion.
Using contour deformation technique for an arbitrary spin $j$
generator we have:
  \ea
\{ W_{-n}\phi_{h_0w_0}(z) \} \v{hw}= \Biggl \{ {{(-1)^{n+j}}
\over {z^n} } \biggl [ {n-1 \choose n-j } w_0 - \sum_{k=1} ^{j-1} z^k
f_{k,n,j}(1)  W_{-k}  \biggr ] \phi_{h_0w_0}(z) \Biggr \}\v{hw}\nm
+ \Biggl \{ {{(-1)^{n+j}}
\over {z^n} } \sum_{k=1} ^{j-1} z^k f_{k,n,j}(q) W_{-k}
   +\sum_{l=0} ^{\infty }{n+l-j \choose l}(pz)^l W_{-n-l}\Biggr \}
   \phi_{h_0w_0}(z)\v{hw}
\label{fuse}
  \eea
 where we have used $ f_{k,n,j}(q)=\sum_{i=1} ^{k } {n-i-1 \choose n-j}
 {j-i-1 \choose k-i}q^{k-i} $.
The fusion point, the argument of the resulting field,
is $qz$ and $p=1-q$.
For $j=2$ the  formula yields the result of \cite{Bau} if we use
$L_{-1}\phi(z)=\partial_z \phi(z)$. Unfortunately in the
\bc, $j=4 $, case we have unknown modes: $W_{-1}\ ; W_{-2}\ ;W_{-3} $.
However following the idea of Bowcock and Watts \cite{BoWa} we are
able to solve this problem. In the $WA_2$ case they could eliminate
this kind of modes considering not only the two independent null
vectors but also their descendants simultaneously.

For simplicity we deal with the following case: completely
degenerate representations having null vectors at level one and two.
First we analyze the $u=\v{21;11}$ field.
The $h,w$ of this field are $\Delta, \omega $, respectively:
\ea
\Delta &=&{1\over 4}(5Q-6)  \nm
\omega &=& -{Q\over 8}\Delta
            \sqrt {{{(4Q-5)({8\over Q}-5)({2\over Q}-3)({6\over Q}-7)}
               \over{(226-75Q-{150\over Q})(Q-3)(3Q-7)}}}
\eea
Analyzing the $C_2$ Toda model in WZNW framework we obtained \cite{Me}
that the only relevant operator, $u$, has a null state at level four.
Studying the covariance of this quantum equation of motion we
concluded that $u$ must belong to a completely degenerate
representation and so it has the following null states.
At level one the null vector is:
\e
{\cal O}_{-1}u=W_{-1}u +\beta^{ 1} L_{-1}u=0
\ee
where
$$
 \beta^{1}=-2(4Q-5)(8/Q-5)(7Q-6)(3Q-2)N
$$
and
$$
 N= {\omega \over \Delta }{1\over{(4Q-5)(8/Q-5)(7Q-6)(3Q-2)}}
$$
At level two this completely degenerate representation has an
independent null vector. However it is more convenient to use the
following null state:
\e
{\cal O}_{-2}u=W_{-2}u+\beta^{ 2}L_{-2}u+\beta^{ 1 1}L_{-1}^2u=0
\ee
$$
 \beta^{2}= 2(23-10Q)(7Q-6)(3Q-2)N \quad ;\quad
 \beta^{11}= 4(13Q-25)(7Q-6)(3Q-2)N
$$
We use the following descendant null vector of the two
independent null vectors at level three:
\e
{\cal O}_{-3}u=W_{-3}u+\beta^{3}L_{-3}u+\beta^{12}L_{-1}L_{-2}u
                        +\beta^{111}L_{-1}^3u=0
\ee
where
$$
\beta^{111}= {16\over Q}(75Q^2-226Q+150)N
$$
$$
 \beta^{3}= -6(5Q-22)(4Q-5)(Q-2)N \quad ;\quad
  \beta^{12}= -24(34Q^2-113Q+82)N
$$
Finally $u$ satisfies the quantum Toda equation of motion,
which is again a descendant null vector:
\e
{\cal O}_{-4}u=W_{-4}u+\beta^{4}L_{-4}u+\beta^{22}L_{-2}^2u
+\beta^{13}L_{-1}L_{-3}u+\beta^{112}L_{-1}^2L_{-2}u
+\beta^{1111}L_{-1}^4u=0
\ee
where
$$
 \beta^{4}=-{2\over Q}(3Q-2)(10Q^3-73Q^2+312Q-360)N
$$
$$
 \beta^{13}=-8(27Q^3-263Q^2+548Q-300)N \quad ;
 \quad \beta^{22}=4(3Q-2)(16Q-27)N
$$
$$
 \beta^{1111}=Q\beta^{111} \quad ;\quad  \beta^{1112}=-\beta^{111}
$$
The first three null vectors carry enough information to
eliminate the unknown $W$ modes while we can use the Toda
equation to fuse with another primary field.
This equation indicates which representations can occur after
the fusion. We did this analysis in \cite{Me} and found
that the non vanishing matrix elements of $u$ are:
\ea
\v{21;11} \times \v{ab;cd} \to \matrix
                  {&&\vert  a+1\  \ b\ ;c\ d \rangle \nm
                   &&\vert  a-1\ \ b \ ;c\ d \rangle \nm
                   &&\vert a+1\ b-1;c\ d \rangle \nm
                   &&\vert a-1\ b+1;c\ d \rangle \nm }
\eea
For producing null vectors we consider the following fusion:
\e
\v{21;11} \times \vert 0r;ss^{'} \rangle \to \vert 1r;ss^{'} \rangle
\ee
Using the method of \cite{Bau} we expand the two OPE as
\e
u(z)\vert 0r;ss^{'} \rangle =\sum_{n\ge 0} z^{n-y}f_n \quad ;
\quad \{{\cal O}_{-4}u(z)\}\vert 0r;ss^{'} \rangle
=\sum_{n\ge 0} z^{n-y-4}\tilde f_n
\ee
where $y=\Delta+h_{0r;ss^{'}}-h_{1r;ss^{'}}$.
Now using (\ref{fuse}) and choosing the fusion point at $p=1\ ;\ q=0 $
we get the following recursive equation:
\e
\tilde f_n =\alpha _n f_n +\sum_{j\ge 1} \biggl \{
 W_{-j}+\gamma_{n,j}L_{-j}+\beta^{22} \Lambda_{-j} \biggr \}
\ee
where
$$
\alpha_n=N(75Q^2-226Q+150){16\over Q^2}n(n-s)(n+rQ-s-s^{'})
(n+rQ-2s-s^{'})
$$
\ea
\gamma_{n,j}=&\beta^{112}(n-y-2)(n-y-3)+\beta^{13}(n-y-3)(j-2)
  +\beta^{4}(j-2)(j-3)/2 \nm &+3\bigl(\beta^{2}+\beta^{12}(n-y-2)
                                 +\beta^{3}(j-2) \bigr)
\eea
and $\Lambda_{-j}=\displaystyle\sum_{i \ge 1}^{j-1}L_{-i}L_{-j+i}$.
The vanishing of $\alpha_n$ at $n=0$ and $n=s$ explicitly shows
that the fusion is allowed and that there is a null
state at level $s$, respectively.
Solving these equations iteratively we find
for the null state at level $s$:
\e
{\cal O}_{1;s}\vert 1r;ss^{'} \rangle =
\sum_{n_i:\sum_{i=1}^p n_i=s}{{\prod_{i=1}^p
  ( W_{-n_i}+\gamma_{N_i,n_i}L_{-n_i}+\beta^{22}\Lambda_{-n_i})}
  \over {\prod_{i=1}^{p-1}\alpha_{N_i}}}\vert 1r;ss^{'} \rangle
 \ee
where $N_i=\sum_{j=1}^i n_j $. Here and from now on we use the
ordering from left to the right.
Alternative expressions can be obtained by choosing the fusion
point in a different way. The result is very elegant if we use
$q=1$. In this case :
\e
{\cal O}_{1;s}\vert 1r;ss^{'} \rangle =
\sum_{n_i=1,2,3,4:\sum_{i=1}^p n_i=s}{{\prod_{i=1}^p
   \Gamma_{n_i}^{N_i}}
  \over {\prod_{i=1}^{p-1}\alpha_{N_i}}}\quad \vert 1r;ss^{'} \rangle
 \ee
where
\ea
\Gamma^n_1&=&{\cal O}_{-1}
+\biggl ((n-y-1)\bigl(6\beta^{11}+9\beta^{111}(n-y-2)
+4\beta^{1111}(n-y-2)(n-y-3)\bigr)\nm
 &+&3\beta^{21}(h+y+2-n)+2\beta^{112}(n-y-3)(h+y-n)
 -\beta^{13}(2h+y+1-n)\biggr ) L_{-1}\nm
 \Gamma^n_2&=&3{\cal O}_{-2}+\biggl((n-y-2)\bigl(3\beta^{12}
   +\beta^{112}(n-y-3)\bigr)+2\beta^{22}(h+y+2-n)\biggr)L_{-2}\nm
&+&\biggl(9\beta^{111}(n-y-2)+6\beta^{1111}
(n-y-2)(n-y-3)+\beta^{112}(h+y+2-n) \biggr )L_{-1}^2 \nm
\Gamma_3^n&=& 3{\cal O}_{-3}+(n-y-3)\bigl(\beta^{13}L_{-3}
+2\beta^{112}L_{-1}L_{-2}+4\beta^{1111}L_{-1}^3\bigr) \nm
\Gamma_4^n&=&{\cal O}_{-4}
\eea
We remark that the computations for the $\v{11;12}$ state proceed in
the same way as in the $\v{21;11}$ case; the final result can be obtained
form eq. (15)-(17) by the $Q \to {2\over Q} $ substitution.
The null vector produced in this manner is in
the $\v{ss^{'};r1}$ representation at level $s$.

There are two other cases when the representation contains two
independent null vectors at level one and two. The first can be
described by  $\tilde u=\vert 12;11 \rangle $ while the other one
can be obtained from this one by the $Q \to {2\over Q} $ substitution.
Both correspond to the Toda field of the $B_2$ Toda model.
 The $h,w$ weights of $\tilde u$ are:
\ea
\tilde \Delta &=&2Q-2 \nm
\tilde \omega &=& {1\over 2}\tilde\Delta
            \sqrt {{{(4Q-5)({2\over Q}-3)({6\over Q}-7)(Q-3)}
               \over{(226-75Q-{150\over Q})({8\over Q}-5)(3Q-7)}}}
\eea
As before, we analyze the leading null vectors of $\tilde u$ and their
descendants simultaneously. The $\tilde u $ field satisfies
equations entirely analogous to eq.(6)-(8) the only difference is that
we have to substitute every $\beta $ with the following $\tilde \beta $:
At level one we have:
$$
\tilde\beta^{1}=-2(4Q-5)(Q-3)(7Q-6)(3Q-2)(Q+1)\tilde N
$$
where
$$
 \tilde N= {\tilde \omega \over \tilde \Delta }
 {1\over{(4Q-5)(Q-3)(7Q-6)(3Q-2)(Q+1)}}
$$
At level two we use again instead of the independent null states their
convenient combination whose coefficients are:
$$
\tilde\beta^{2}=4Q(Q-1)(16Q-27)(Q+1)(3Q-2)\tilde N \quad ; \quad
\tilde\beta^{11}=-(59Q^2-199Q+50)(Q+1)(3Q-2)\tilde N
$$
The coefficients of the null states at level three are:
$\tilde\beta^{12}=6Q(41Q^2-113Q+68)(Q+1)\tilde N$
$$
\tilde\beta^{3}=6Q(7Q-6)(Q-3)^2(Q+1)\tilde N \qquad ; \quad
\tilde\beta^{111}={Q(Q+1)\over 8}\beta^{111}
$$
However, contrary to the previous case, this representation has no null
states of the form of (9) at level four. Analyzing the $B_2$ Toda
model we learn that $\tilde u$ has to satisfy the following equation,
which shows the existence of a descendant null state at level five:
\ea
&{\cal O}_{-5}\tilde u=
\tilde\beta^{14}L_{-1}L_{-4}u-{2\over (Q+1)}\tilde\beta L_{-1}W_{-4}u
+\tilde\beta W_{-5}u+\tilde\beta^{122}L_{-1}L_{-2}^2u
+\tilde\beta^{113}L_{-1}^2L_{-3}u \nm
 &\qquad +\tilde\beta^{1112}L_{-1}^3L_{-2}u
+\tilde\beta^{11111}L_{-1}^5u+\tilde\beta^{23}L_{-2}L_{-3}u
+\tilde\beta^{5}L_{-5}u=0 \hskip 2cm
\eea
where
$$
\tilde\beta^{5}=4(Q+1)(Q-3)(3Q^3-2Q^2+48Q-60)\tilde N \quad ; \quad
\tilde\beta^{1112}=-{Q\over 4}\beta^{111}
$$
$$
\tilde\beta^{14}=-6(Q-3)(3Q^3-2Q^2+48Q-60)\tilde N \quad ; \quad
\tilde\beta^{113}=-6(Q-3)(7Q^2-54Q+50)\tilde N
$$
$$
\tilde\beta^{122}=2Q(Q-3)(27Q-32)\tilde N \quad ; \quad
\tilde\beta^{23}= -2Q(Q-3)(Q+1)(27Q-32)\tilde N \quad ; \quad
\tilde\beta^{11111}={1\over 8}\beta^{111}
$$
Eq.(19) is the quantum equation of motion of the $B_2$ Toda model.
{}From this equation we conclude that the possible fusions of
$\tilde u$ are the following:
\ea
\v{12;11} \times \v{a\ b;c\ d} \to
           \matrix {&&\v{ a\  b+1;c\ d} \nm
                   &&\v{ a-2\ b+1;c\ d } \nm
                   &&\v{ a\  b ;c\ d } \nm
          &&\v{ a\ b-1;c\ d }& \nm
          &&\v{ a+2\ b-1;c\ d } \nm  }
\eea
Considering the
\e
\v{12;11} \times \vert r0;ss^{'} \rangle \to \vert r1;ss^{'} \rangle
\ee
fusion and using the same technique as before: i.e. expanding the
two OPE as
\e
u(z)\vert r0;ss^{'} \rangle =\sum_{n\ge 0} z^{n-y}f_n \quad ;\quad
\{ {\cal O}_{-5}u(z)\} \vert r0;ss^{'} \rangle
= \sum_{n\ge 0} z^{n-y-5}\tilde f_n
\ee
with $y=\tilde\Delta+h_{r0;ss^{'}}-h_{r1;ss^{'}}$
and using (4) we obtain
\e
\tilde f_n =\tilde\alpha _n f_n +\sum_{j\ge 1} \biggl \{
 \tilde N(j-4-{2(n-y-4)\over {Q+1}})W_{-j}
 +\tilde\gamma_{n,j}L_{-j}+\tilde\beta^{122}(n-y-4) \Lambda_{-j}
 +\tilde\beta^{23}\Lambda^{'}_{-j} \biggr \}
\ee
where
$$
\tilde \alpha_n=\tilde N{2\over Q}(75Q^2-226Q+150)n(n-s^{'})
           (n+{r+1\over 2}Q-s-s^{'})(n+rQ-2s-s^{'})(n+rQ-2s-2s^{'})
$$
and
\ea
&\tilde\gamma_{n,j}=(n-y-4)\biggl(\tilde\beta^{114}{j-2 \choose 2}+
                    (n-y-3)\Bigl(\tilde\beta^{1112}(n-y-4)
                     +\tilde\beta^{113}(j-2)\Bigr)\biggr)\hskip
1cm \nm
&-\tilde\beta^{5}{j-2\choose 3}+2\tilde\beta^{122}(h+y+1-n)
             +\tilde\beta^{23}\bigl((h+3+y-n)(j-2)+n-y-2h)\nm
    &  -{2(n-y-4)\over Q+1 }\bigl(3\tilde\beta^{2}+\tilde\beta^{12}(x-2)
                                 +3\tilde\beta^{3}(j-2) \bigr)
                    - \bigl(8\tilde\beta^{2}+\tilde\beta^{12}(x-2)
                       +6\tilde\beta^{3}(j-2) \bigr);
\eea
 $\Lambda_{-j} $ is the same as before and  $\Lambda^{'}_{-j}
  =\sum_{i \ge 1}^{j-1}(i-2)L_{-i}L_{-j+i}$.
One can prove that in the $\v{r1;ss^{'}} $ representation  there is a
null state at level $s^{'} $.
An explicit expression for this null vector can be obtained by solving
the recursion:
\ea
{\cal O}_{1;s^{'}}\vert r1;ss^{'} \rangle = \hskip 14cm \nm
\lefteqn=\displaystyle\sum_{n_i:\sum_{i=1}^p n_i=s}{{\prod_{i=1}^p
\biggl \{  \tilde N(j-4-{2(n-y-4)\over {Q+1}})W_{-j}
 +\tilde\gamma_{n,j}L_{-j}+\tilde\beta^{122}(n-y-4) \Lambda_{-j}
 +\tilde\beta^{23}\Lambda^{'}_{-j} \biggr \}  }
  \over {\prod_{i=1}^{p-1}\tilde\alpha_{N_i}}}\vert r1;ss^{'} \rangle
\nonumber
\eea
where $N_i=\sum_{j=1}^i n_j $.
An alternative expression can be obtained by choosing the fusion
point at $q=1$, however this straightforward formula looks complicated
so we omit it here.

Summarizing we find that the fusion rules of the Toda fields
--$u, \tilde u $ -- are the same as the fusion rules of the
$(10),\  (01)$ representations of the corresponding Lie algebra.
However the fusions of the various Toda fields are determined by the
quantum equations of motion of the corresponding Toda models
and these quantum equations of motion take the form of particular
descendant null vectors. So we conclude that for producing null
vectors via the fusion method one has to consider these descendant
null vectors. Furthermore we conjecture how to generate null vectors
in the most general case, i.e. in the highest weight representation
of the \bc algebra corresponding to the $\v{rr^{'},ss^{'}}$ hw. vector.
One has to consider the WZNW model associated not to the defining but
to a higher dimensional representation, labelled by $(r,0)$,
of the appropriate group and use the Drinfeld-Sokolov reduction
\cite{DS} to obtain the
corresponding Toda model. Fusing its $\v{r1,11}$ type special
descendant null vector with a $\v{0r^{'},ss^{'}}$ primary field
a null vector of the above mentioned representation is hoped to be
obtained.

\smallskip

I would like to thank L. Palla and P. B\'antay for the many
useful discussions and comments and L. P. for the careful
reading of the manuscript.

\end{document}